\documentclass[aps,prb,twocolumn,superscriptaddress,showpacs]{revtex4}
\usepackage{graphicx}
\usepackage{amsmath,amssymb}
\usepackage{ulem}

\usepackage[colorlinks]{hyperref}
\hypersetup{
pdfstartview=Fit,
pdfpagemode=UseOutlines,
pdfdisplaydoctitle=true,
pdfauthor={B. Nafradi et al.},
pdftitle={EDT-AsF6 frustration},
colorlinks=true,
citecolor={blue},
linkcolor={blue},
urlcolor={black},
breaklinks=true,
bookmarksopen=true,
bookmarksopenlevel=5,
bookmarksnumbered=true
}

\begin{document}

\title{Frustration induced one-dimensionality in the isosceles triangular antiferromagnetic lattice of $\delta$-(EDT-TTF-CONMe$_{2}$)$_{2}$AsF$_6$}
\author{B.~N\'afr\'adi}
\affiliation{Institute of Physics, Ecole Polytechnique F\'ed\'erale de Lausanne (EPFL), CH-1015 Lausanne, Switzerland}
\affiliation{Department of Physics, Budapest University of Technology and Economics, Budafoki \'ut 8., H-1111 Budapest, Hungary}
\author{\'A.~Antal}
\affiliation{Department of Physics, Budapest University of Technology and Economics, Budafoki \'ut 8., H-1111 Budapest, Hungary}
\affiliation{Institute of Physics, Ecole Polytechnique F\'ed\'erale de Lausanne (EPFL), CH-1015 Lausanne, Switzerland}
\author{T.~Feh\'er}
\affiliation{Department of Physics, Budapest University of Technology and Economics and MTA-BME Condensed Matter Research Group and MTA-BME Lend\"ulet Magneto-optical Research Group and Spintronics Research Group, Budafoki \'ut 8., H-1111 Budapest, Hungary}
\author{L.~F.~Kiss}
\affiliation{Wigner Research Centre for Physics of the Hungarian Academy of Sciences, H-1525 Budapest, Hungary}
\author{C.~M\'ezi\`ere}
\affiliation{Laboratory MOLTECH-Anjou, University of Angers, CNRS UMR 6200, 49045 Angers, France}
\author{P.~Batail}
\affiliation{Laboratory MOLTECH-Anjou, University of Angers, CNRS UMR 6200, 49045 Angers, France}
\author{L.~Forr\'o}
\affiliation{Institute of Physics, Ecole Polytechnique F\'ed\'erale de Lausanne (EPFL), CH-1015 Lausanne, Switzerland}
\author{A.~J\'anossy}
\affiliation{Department of Physics, Budapest University of Technology and Economics, Budafoki \'ut 8., H-1111 Budapest, Hungary}

\date{\today}

\begin{abstract}

The $1/4$-filled  organic compound, $\delta$-(EDT-TTF-CONMe$_{2}$)$_{2}$AsF$_6$ is a frustrated two-dimensional triangular magnetic system as shown by high-frequency (111.2 and 222.4~GHz) electron spin resonance (ESR) and structural data in the literature. 
The material gradually orders antiferromagnetically below 40~K but some magnetically disordered domains persist down to 4~K. 
We propose that in defect free regions frustration prevents true magnetic order down to at least 4~K in spite of the large first- and second-neighbor exchange interactions along chains and between chains, respectively. 
The antiferromagnetic (AFM) order gradually developing below 40~K nucleates around structural defects that locally cancel frustration. 
Two antiferromagnetic resonance modes mapped in the principal planes at 4~K are assigned to the very weakly interacting one-dimensional molecular chains in antiferromagnetic regions.
\end{abstract}


\maketitle

\section{Introduction}

Materials with dense magnetic atomic or molecular sites usually have a magnetically ordered ground state.
However, in some two-dimensional (2D) lattices quantum fluctuations and frustration of interactions between neighboring magnetic sites preclude long range order. In these the rotational symmetry is preserved and the ground state is an exotic spin liquid \cite{Fazekas1974,Anderson1987}.
Kagome lattices with isotropic Heisenberg interactions between sites are prime examples \cite{Balents2010}.
However, it was noticed in the experimental realizations, vesignieite \cite{Okamoto2009} and herbertsmithite, \cite{Olariu2008} that in the complex kagome geometry a strong Dzyaloshinskii-Moriya (DM) interaction freezes the quantum spin liquid at finite temperatures \cite{Zorko2013}. 
A simpler configuration is a triangular lattice of spin-1/2 moments with isotropic antiferromagnetic Heisenberg exchange interaction between sites. 
The equilateral triangular lattice with classical spins has an ordered ground state, but  frustration in the quantum spin lattice leads to a spin liquid which is readily studied in organic materials like $\kappa$-(BEDT-TTF)$_2$Cu$_2$(CN)$_3$ and EtMe$_3$Sb[Pb(dmit)$_2$]$_2$ \cite{Kanoda2011}. 

A question of great interest is the robustness of the spin liquid state against deviations from the ideal structure. 
Namely, the nature of the magnetic ground state of the isosceles triangular antiferromagnetic lattice with different first-neighbor interactions, $J$ and $J_2$, on the base and legs respectively, is strongly debated \cite{Chung2001,Scriven2012,Reuther2011}. 
It is believed that the system decouples into weakly interacting antiferromagnetic 1D chains if $J$ is much larger than $J_2$ and a spin liquid state is formed with no magnetic order. 
The inorganic materials Cs$_2$CuCl$_4$ and Cs$_2$CuBr$_4$ were studied within this context \cite{Zvyagin2014}.
In these systems the ground state is extremely sensitive to magnetic fields due to deviations from the isotropic exchange Heisenberg model, in particular the DM interaction \cite{Starykh2010, Povarov2011,Zorko2013}.

The organic magnet $\delta$-(EDT-TTF-CONMe$_{2}$)$_{2}$AsF$_6$, hereafter EDT$_2$AsF$_6$, \footnote{$\delta$-(EDT-TTF-CONMe$_{2}$)$_{2}$AsF$_6$, where EDT stands for the tertiary amide-functionalized ethylenedithiotetrathiafulvalene} is an excellent model system for studying frustration in an isosceles triangular lattice.
The asymmetric EDT molecules (Fig.~\ref{Fig.Structure1}~(A), further on symbolized by a duck) form chains.
The molecular separation is uniform along chains, the instability of the quasi-1D electronic system results in alternating ``charge rich'' and ``charge poor'' molecules with charges of 0.9~e$^+$ and 0.1~e$^+$, respectively \cite{Heuze2003,Auban2009,Zorina2009} (dark blue and light blue ducks in Fig.~\ref{Fig.Structure1}~(B)).
The magnetic structure sketched in Fig.~\ref{Fig.Structure1}~(C) is approximated by a two-dimensional triangular system, with the exchange interactions $J$, $J_2$ and $J_3$.
The high-temperature magnetic properties follow a 1D spin-$1/2$ Heisenberg antiferromagnetic chain with an isotropic exchange of $J=298$~K \cite{Nafradi2010}.
There is no direct measurement of $J_2$; we estimate it from the calculated overlap integrals along and perpendicular to the chains \cite{Heuze2003} to be about 30~K. 
Overlap is small in the \textbf{c} direction where AsF$_6$ ions separate the chains.
The exchange parameters are somewhat modified below the orthorhombic to monoclinic transition \cite{Zorina2009} at 190~K. 
Anisotropies, deviations from the isotropic exchange Heisenberg model are weak \cite{Nafradi2010}.
An antiferromagnetic (AF) ordering has been observed\cite{Auban2009} at T$_ \text{N}=8.5~K$.  
The DM interaction between molecules along chains in the \textbf{a} direction is forbidden by symmetry \cite{Zorina2009} in the AF state and the first neighbor dipole-dipole interaction is also canceled (see \ref{comparison}).

Here we report high-frequency electron spin resonance (ESR) measurements (complemented by static magnetization data) which elucidate the magnetic structure in EDT$_2$AsF$_6$.
The magnetic resonances of the antiferromagnetically ordered and the paramagnetic states are well separated in the ESR spectrum.
Some paramagnetic regions persist to temperatures as low as 4~K.
This is unexpected since the overlap integral between molecules on adjacent chains along \textbf{b} is quite large \cite{Heuze2003} and one expects intuitively a long range ordered AFM state to set in at much higher temperatures.
We suggest that EDT$_2$AsF$_6$ represents the remarkable case of an isosceles triangular lattice which is transformed by frustration into a system of weakly interacting antiferromagnetic chains.
Without defects there would be no magnetic order down to the temperature range we investigated.
However, defects locally lift the frustration in the imperfect crystals.
As a result, at 4~K most of the crystal is antiferromagnetically ordered and depending on the crystal quality, some antiferromagnetic regions persist up to 40~K.
The ESR in antiferromagnetically ordered regions shows that the chains (along \textbf{a}) interact surprisingly weakly in the \textbf{b} direction.
This again is a result of frustration canceling the interaction between chains.

\section{Experimental}
EDT$_2$AsF$_6$ single crystals were grown using electrochemical oxidation of EDT like in earlier studies of the same compound \cite{Heuze2003,Auban2009,Zorina2009,Nafradi2010}.
ESR was performed by home-built spectrometers operating at 111.2 and 222.4~GHz frequencies \cite{Nafradi2008a,Nafradi2008b,Nagy2011}.
These setups are particularly suitable to detect the AFM resonance in organic magnets.
The ESR spectra presented in this paper were recorded at fixed frequencies sweeping the magnetic field and measuring the derivative of the microwave intensity reflected from the sample.

The static magnetic susceptibility measured by SQUID and the paramagnetic susceptibility measured by the ESR intensity were compared in Refs.~\onlinecite{Heuze2003,Nafradi2010}.
The high temperature susceptibility data were used to calibrate the absolute value in our measurement.

\section{Results and Discussion}
\subsection{Paramagnetism and antferromagnetic ordering}

\begin{figure}
	\includegraphics[width=7.9cm]{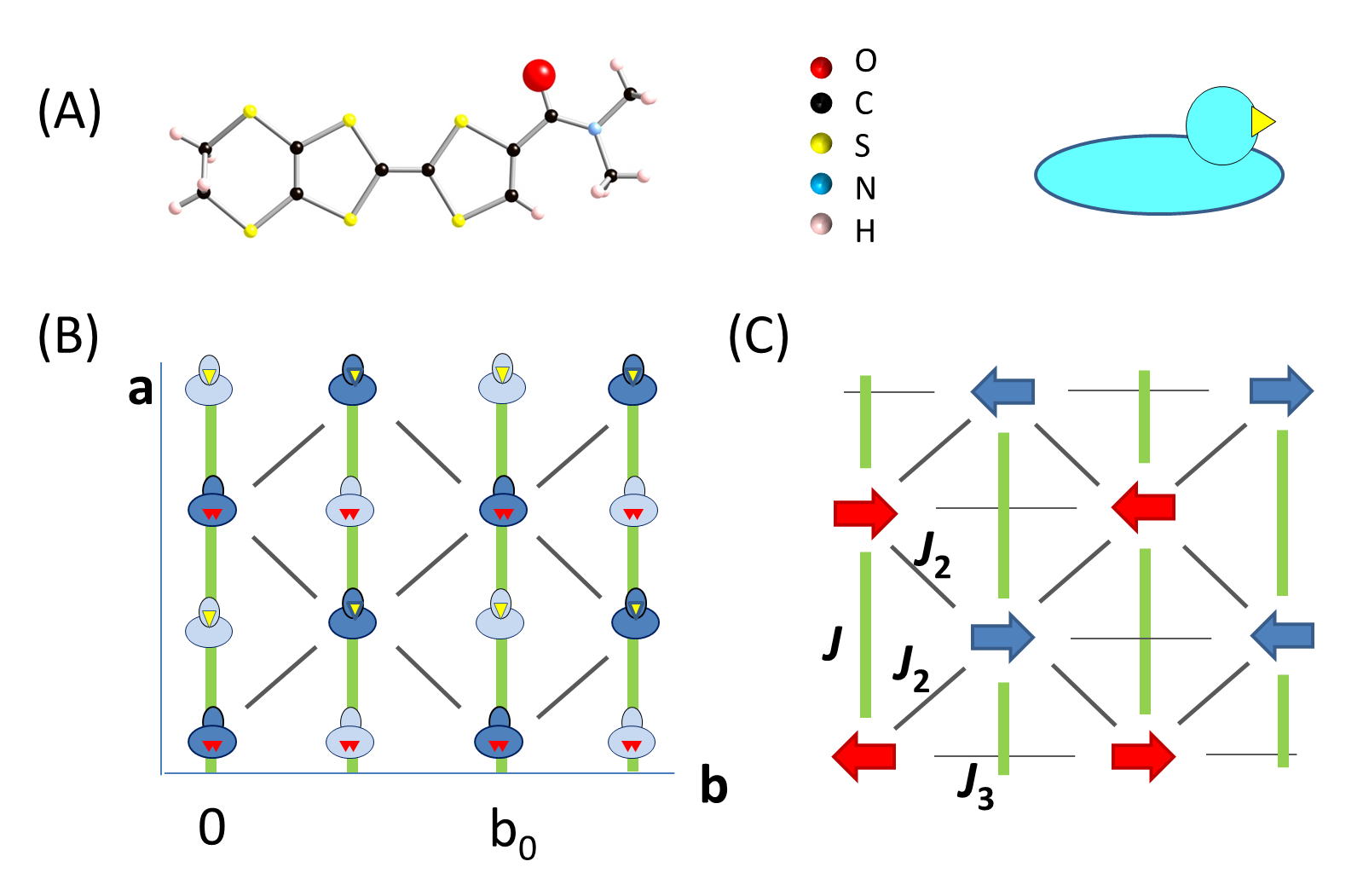}
   \caption{(A) The EDT molecule and its schematic representation as a duck in further panels. (B) Schematic structure of the (\textbf{a},\textbf{b}) plane of EDT$_2$AsF$_6$. Dark and light ducks represent charge rich and charge poor molecules. (C) The frustrated, isosceles triangular magnetic lattice. $J$, $J_2$ and $J_3$ are the first, second and third neighbor isotropic antiferromagnetic exchange interactions respectively. The exchange is large along chains while frustration cancels the isotropic interactions between chains.  \label{Fig.Structure1}}
\end{figure}

\begin{figure}
	\includegraphics[width=7.9cm]{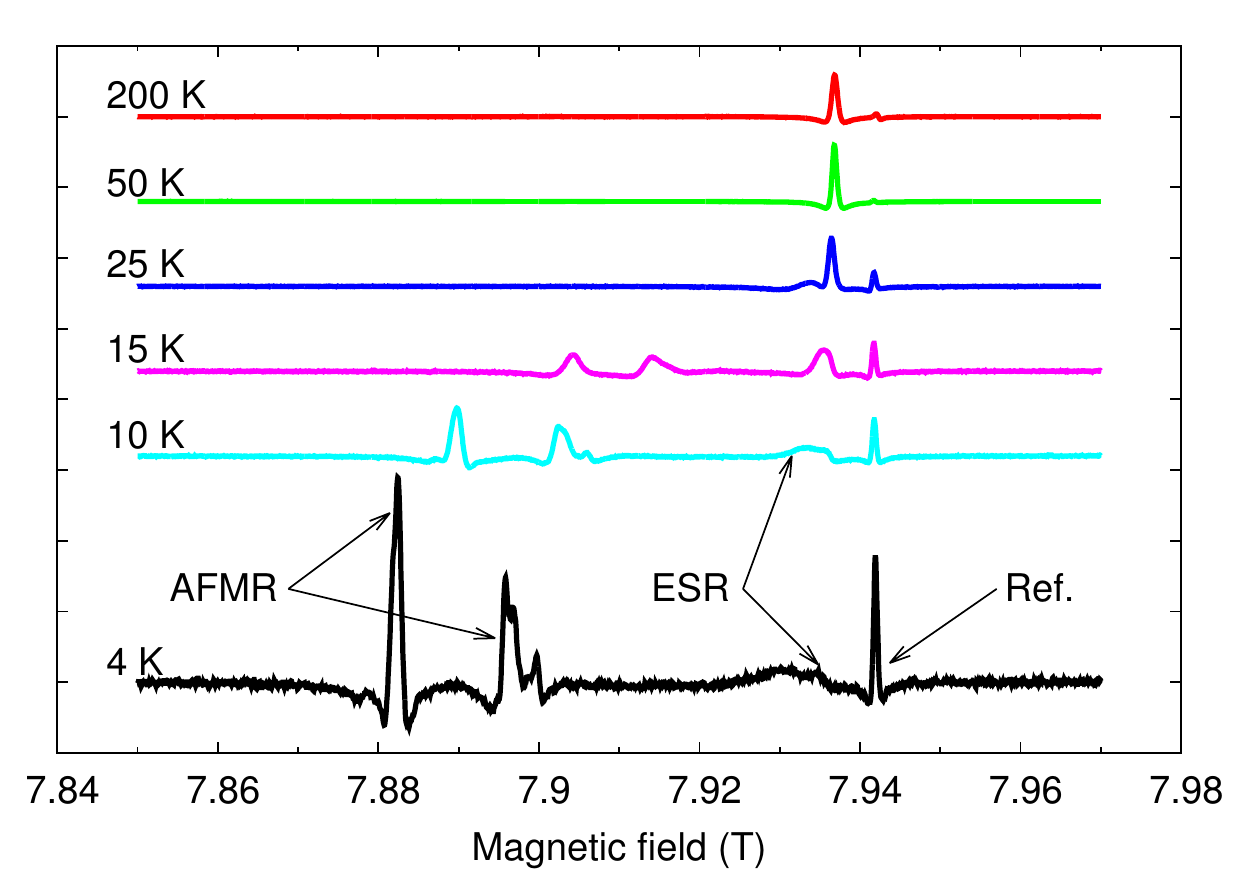}
   \caption{Temperature dependence of the ESR and AFMR spectra in an EDT$_2$AsF$_6$ crystal with $B_0 \parallel \mathbf{a}$, 222.4~GHz. The AFMR below 25~K signals an antiferromagnetic order in large regions. The ESR from paramagnetic regions persists to low temperatures. Reference at 7.941~T is KC$_{60}$.\label{Fig.TN}}
\end{figure}

\begin{figure}
	\includegraphics[width=7.9cm]{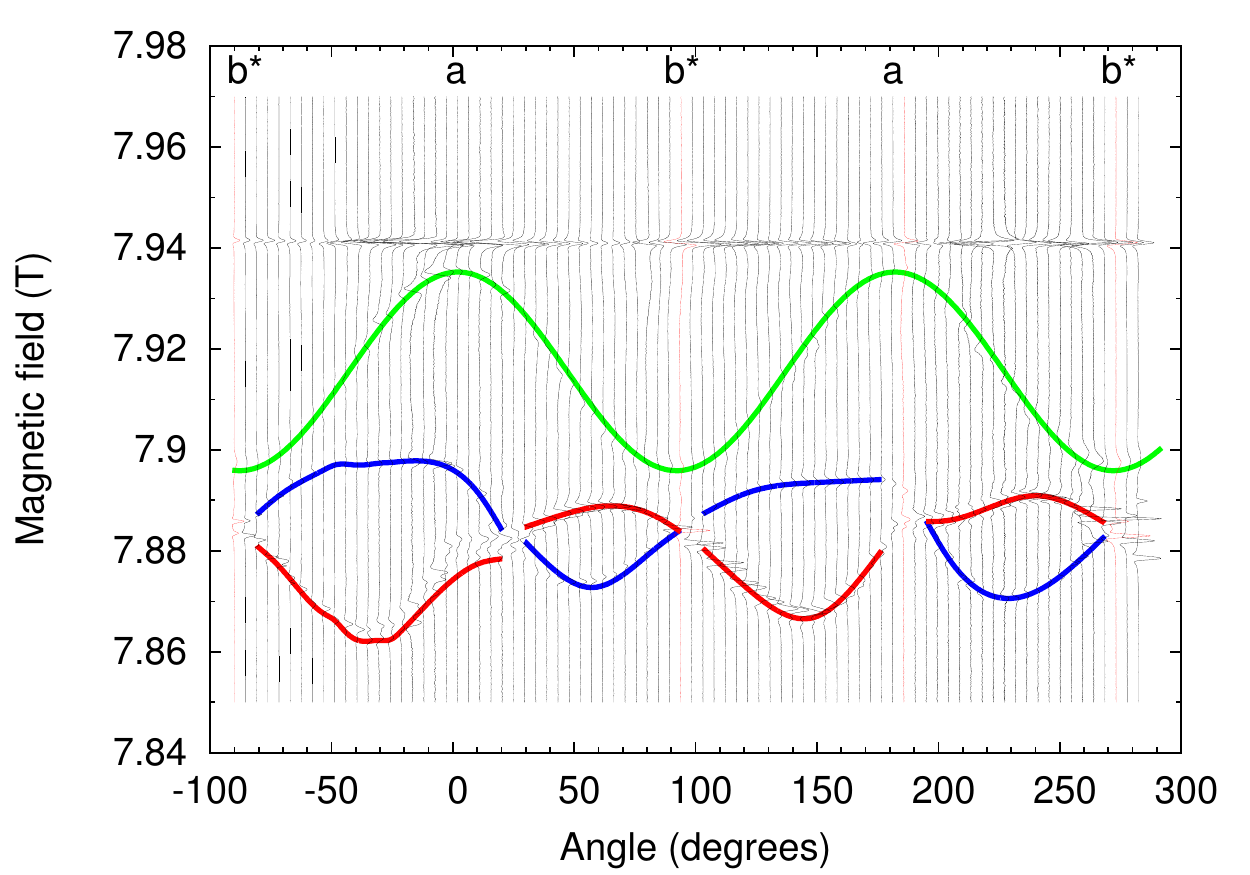}
   \caption{Vertical lines: ESR and AFMR spectra in EDT$_2$AsF$_6$ at 4~K and 222.4~GHz with magnetic field in the (\textbf{a},\textbf{b}*) plane. Red and blue lines: guides to the eye for the AFMR positions.  Green line: ESR resonance fields at 200~K. The ESR resonance field oscillates sinusoidally with angle around 7.92~T. The ESR at 4~K, where most of the sample is antiferromagnetic, has the same g-factor anisotropy as the ESR at 200~K where all of the sample is paramagnetic. The AFMR modes oscillate with magnetic field angle around 7.88~T and  depend on magnetic history. Resonance at 7.941~T is a KC$_{60}$ field reference. \label{Fig.AFMR.ab}}
\end{figure}

Between 40 and 300~K, the static susceptibility follows well the ESR intensity measured at 9~GHz, as expected for a paramagnetic material where spins in the whole crystal contribute to the ESR at $g \approx 2$.
However, below 40~K, the 9~GHz ESR intensity decreases rapidly with decreasing temperature due to the decrease of the paramagnetic regions in the sample \cite{Heuze2003,Nafradi2010}.
The static susceptibility remains large \cite{Heuze2003}, suggesting that below 40~K the material is inhomogeneous; it consists of paramagnetic and antiferromagnetically ordered regions.
These latter ones cannot be detected at 9~GHz since this frequency is within the gap of the excitation spectrum.
At low temperatures a small paramagnetic contribution proportional to the inverse temperature was found.
At 1.8~K the large majority of the powder sample is antiferromagnetic.
At high fields the small paramagnetic term is saturated and the magnetisation, $M$ increases linearly with field.
The effective antiferromagnetic exchange interaction determined from the slope of the magnetization vs. field curve at 1.8~K is roughly $J=410$~K.

The ESR of the paramagnetic material was detected between 4 and 300~K at high excitation frequencies, $ \omega / 2\pi $  = 111.2 and 222.4~GHz. 
The paramagnetic resonance field is proportional to the exciting frequency and has the same $g$-factor anisotropy at all temperatures. 
The resonance splits at low temperatures into the ESR of the paramagnetic domains  and the antiferromagnetic resonance (AFMR) of magnetically ordered domains (Fig.~\ref{Fig.TN}). 
The onset temperature at which the AFMR lines appear resolved from the ESR varies from crystal to crystal between 15 and 40~K. 
The AFMR field is not proportional to the frequency and depends on the anisotropic coupling between sublattice magnetizations.
The AFMR shift from the $g=2$ position at fixed temperature is approximately inversely proportional to $ \omega $ and increases with the increase of the sublattice magnetization at lower temperatures.

The ratio of the ESR and AFMR line intensities decreases gradually below the onset temperature showing the gradual expansion of antiferromagnetic domains at the expense of paramagnetic ones.
However, some regions remain paramagnetic at 4~K since a weak, broadened ESR of uncorrelated EDT$_2$AsF$_6$ chains is still observable.
This ESR line is assigned to paramagnetic EDT$_2$AsF$_6$ regions with few structural defects, embedded in antiferromagnetically ordered parts of the crystal. 
The ESR at 4~K appearing together with the AFMR of most of the sample does not arise from paramagnetic impurities since it has the same g-factor anisotropy as the ESR of the full sample at high temperatures (Fig.~\ref{Fig.AFMR.ab}).
The g factor depends on the orientation and type of the ESR active molecules. 
At the high fields of our experiment, the g-factor anisotropy of an impurity phase would be well resolved from that of paramagnetic EDT$_2$AsF$_6$.

\subsection{Angular dependence of the AFMR modes}

\begin{figure}
	\includegraphics[width=7.9cm]{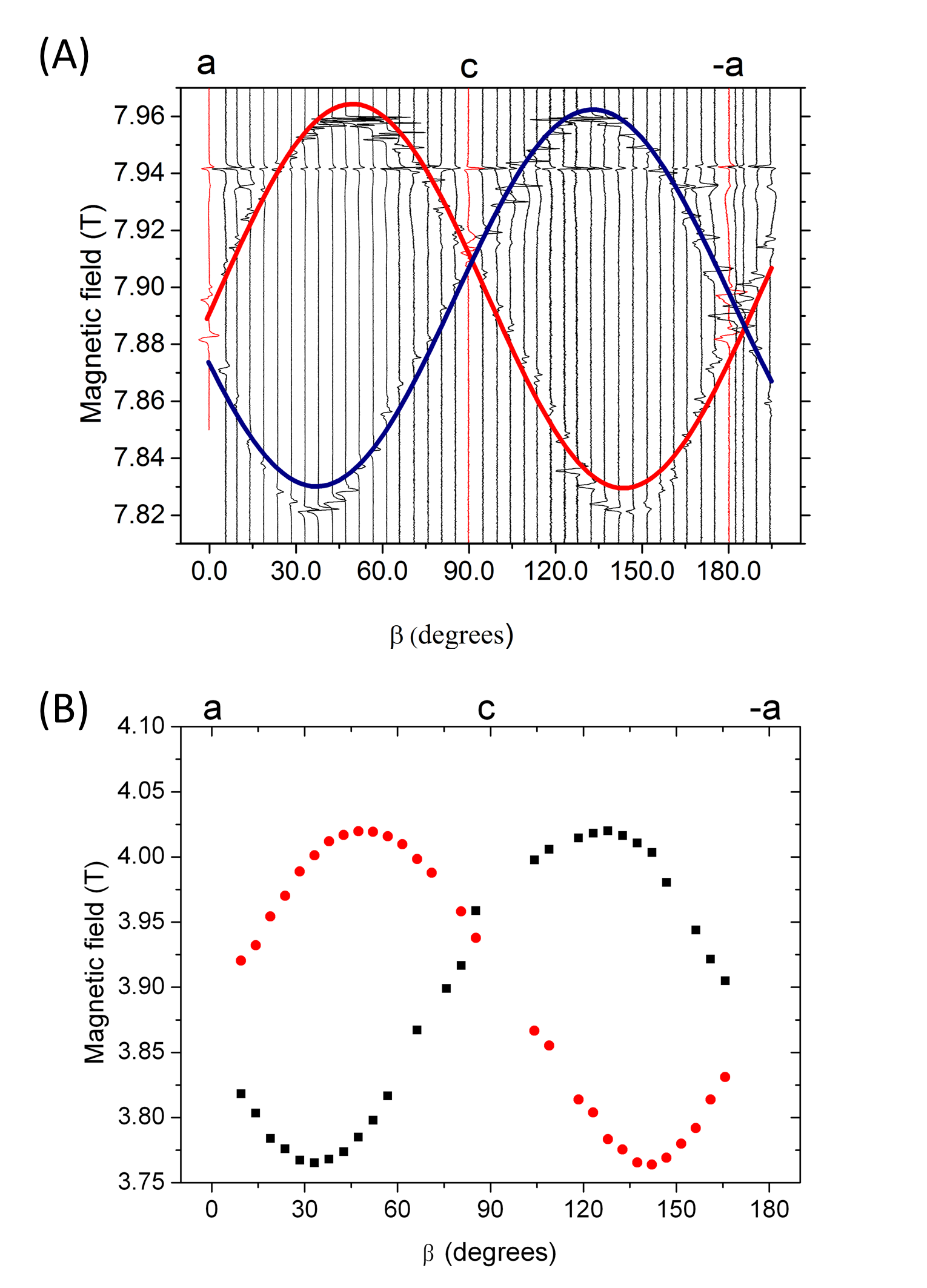}
   \caption{ AFMR modes in EDT$_2$AsF$_6$ with magnetic field in the (\textbf{a},\textbf{c}) plane at 4~K. (A) resonance frquency 222.4~GHz. (Vertical lines are spectra. Red and blue lines are guides to the eye). For $B$ in the (\textbf{a},\textbf{c}) plane, one AFMR mode arises from chains in the b=0 plane, the other from chains in the b=b$_0$/2 planes; (B) The (\textbf{a},\textbf{c}) plane angular dependence of the AFMR resonance field at 111.2~GHz.\label{Fig.AFMR.ac}}
\end{figure}


\begin{figure}
	\includegraphics[width=7.9cm]{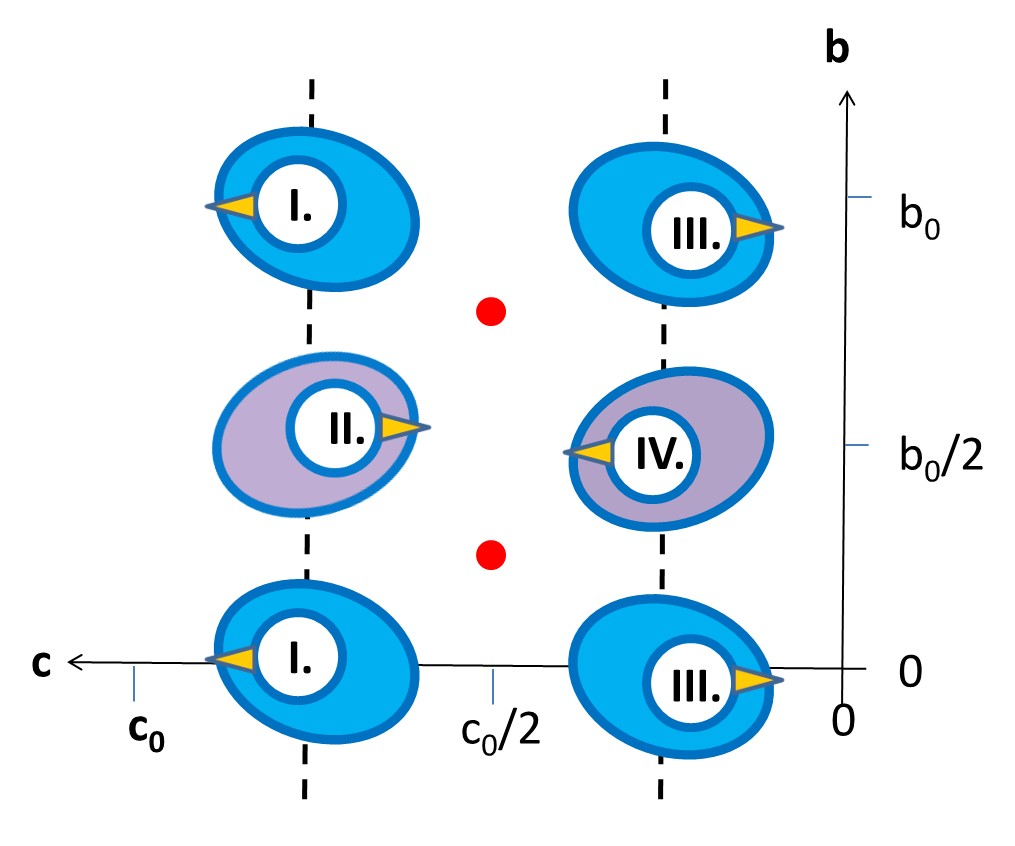}
 \caption{Schematic projection of the (\textbf{b},\textbf{c}) plane along the \textbf{a} (chain) axis showing charged molecules only. Molecules II. and IV. are in the $a=a_0/2$ plane, molecules I. and III. are in the $a=0$ plane.  The AFMR modes are attributed to magnetically weakly interacting chains. For fields in the (\textbf{a},\textbf{c}) plane, the angular dependence of the modes of chains I. and IV. rotate in opposite sense to the modes of chains II. and III. as expected from the glide plane symmetry relation $\textbf{c}\rightarrow-\textbf{c}$ (dashed lines) between the respective chains. The unexpectedly weak effective interaction along \textbf{b} is attributed to magnetic frustration. Interaction is weak in the \textbf{c} direction where AsF$_6^-$ ions (red dots) separate the chains. \label{Fig.AFMR.ac:structure}}
\end{figure}

\begin{figure}
	\includegraphics[width=7.9cm]{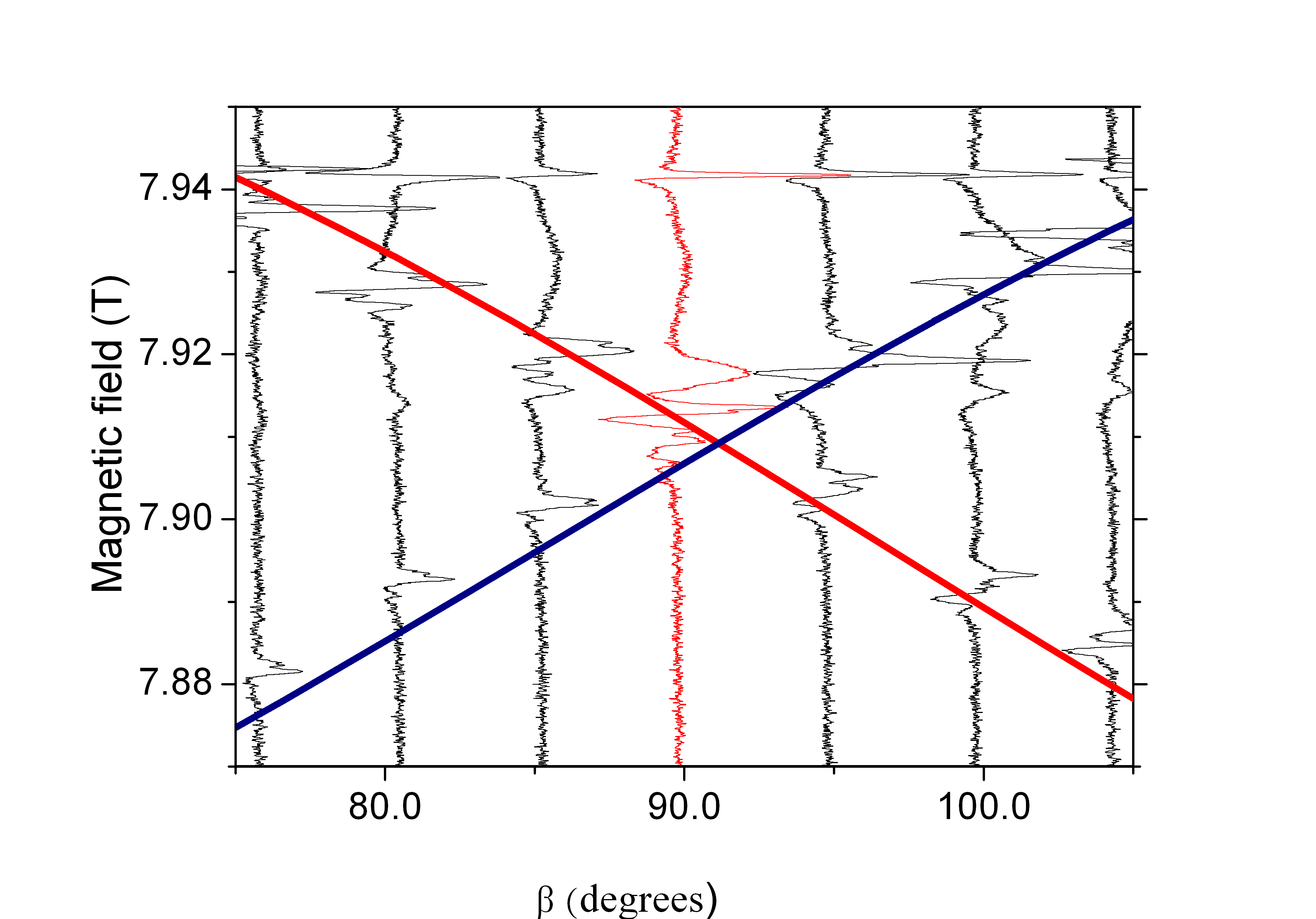}
   \caption{ AFMR in EDT$_2$AsF$_6$ with magnetic field in the (\textbf{a},\textbf{c}) plane, 222.4~GHz in the viciniy of the \textbf{c} direction. ESR line at 7.491~T is a KC$_{60}$ reference.\label{Fig.AFMR.ac:zoom}}
\end{figure}

The angular dependence of the AFMR resonance fields was mapped at 4~K in the (\textbf{a},\textbf{b}*) (Fig.~\ref{Fig.AFMR.ab}), (\textbf{a},\textbf{c}) (Fig.~\ref{Fig.AFMR.ac}~A) and (\textbf{b}*,\textbf{c}) planes at 111.2 and 222.4~GHz.
(\textbf{b}* is perpendicular to the (\textbf{a},\textbf{c}) plane, and is close to \textbf{b}; it coincides with a well-defined edge of the crystal. See § \ref{assignment} for details of the crystal structure).
The accuracy of crystal alignment was better than $5^\circ$.
Two AFMR modes were resolved in general magnetic field directions.
The two modes are degenerate in the \textbf{b}* and \textbf{c} directions but are split in \textbf{a} (Fig.~\ref{Fig.AFMR.ab}). Each mode consists of several closely spaced lines.

In the (\textbf{a},\textbf{b}*) plane, the modes depend on magnetic field history; the hysteresis is strongest near \textbf{a} (see Fig.~\ref{Fig.AFMR.ab}).
We note that line positions are different in subsequent $180^\circ$ rotations of the magnetic field.
The non-symmetric angular dependence around \textbf{b}* is also due to hysteresis.
In our interpretation, it arises from pinning of magnetically ordered domains to a small concentration of defects.

The angular dependence in the (\textbf{b}*,\textbf{c}) plane (not shown) is small and the measured AFMR felds depend sensitively on the precision of the sample orientation.

\subsection{Assignment of AFMR modes} \label{assignment}

We draw our main conclusions from an analysis of the modes measured with field in the (\textbf{a},\textbf{c}) plane (see Fig.~\ref{Fig.AFMR.ac}).
The AFMR fields of the two modes in this plane, $B_{\mathrm{AF}}^{+}$ and $B_{\mathrm{AF}}^{-}$ (red and blue lines, respectively) vary similarly but in opposite sense with the angle $ \beta $ measured from the \textbf{a} axis, i.e. $B_{\mathrm{AF}}^{+}( \beta )$ = $B_{\mathrm{AF}}^{-}(- \beta )$.
Except for the relatively small splitting and hysteresis of lines near \textbf{a}, the angular dependence of the two modes fit well the expressions:

\begin{eqnarray}
B_{\mathrm{AF}}^{+}=B_0+(1/2)b_{ac}\cos(2(\beta_0 + \beta)), \label{eq1} \\
 B_{\mathrm{AF}}^{-}=B_0+(1/2)b_{ac}\cos(2(\beta_0 - \beta)).  \label{eq2} 
\end{eqnarray}

Each curve corresponds to a conventional two-sublattice AFMR excited at frequencies much larger than the gap. 
This agrees with the observation that for the two frequencies, 111.2 and 222.4~GHz, the average  $B_0$ is proportional to $\omega$, and the amplitude, $b_{ac}$ is inversely proportional to $\omega$. 

We argue that the two modes described by Eqs.~\ref{eq1} and \ref{eq2} have the symmetry in the angular dependence $B_{\mathrm{AF}}^{+}( \beta )$ = $B_{\mathrm{AF}}^{-}(- \beta )$
only if the chains are weakly coupled. The inevitable splitting of the modes near the apparent mode crossings is smaller than the experimental uncertainities. 
We assign the two AFMR modes to nearly independent modes of the chemically identical but structurally non-equivalent chains along \textbf{a}.

To explain, we first recall some details of the crystal structure \cite{Heuze2003,Auban2009,Zorina2009}. The full structure including the CDW wavevectors has been determined at room temperatures; NMR shows that the CDW changes little at lower temperatures. Above 190~K the structure is orthorhombic.  A monoclinic distortion arises below 190~K that gradually increases the angle $ \gamma$ from 90 to about 93$^\circ$ at 100~K where it is close to saturation.  This small distortion is assumed to be unimportant and we base the argument on the orthorhombic structure where not stated otherwise. Crystals are twinned, the twins are related by a reflection of all three principal axes above 190~K, thus they are magnetically equivalent in the sense that their ESR spectra are the same. Below 190~K the reflection symmetry between the \textbf{b} axes of the twins is broken, while it is unchanged for the \textbf{a} and \textbf{c} axes. Twins may split somewhat the observed AFMR spectra in general directions but not in magnetic fields in the (\textbf{a},\textbf{c}) plane. 

There are 4 chemically equivalent ``\textbf{a}'' chains that are structurally related by symmetry operations (Fig.~\ref{Fig.AFMR.ac:structure}). Chains I. and III. are in the $b=0$ plane while chains II. and IV. in the  $b=b_0/2$ plane.  Chains I. and III. (similarly II. and IV.) are related by the rotation $[x,y,z]\rightarrow [x,-y,-z]$. Chains I. and II. (III. and IV.) are related by the glide reflection $[x,y,z]\rightarrow [\tfrac{1}{2}+x,\tfrac{1}{2}+y,\tfrac{1}{2}-z]$. Finally, the glide reflection $[x,y,z]\rightarrow [\tfrac{1}{2}+x,\tfrac{1}{2}-y,\tfrac{1}{2}+z]$ relates chains I. and IV. (II. and III.). 

Clearly, if chains did not interact with their neighbors, the reflection symmetry $\textbf{b}\rightarrow -\textbf{b}$ would ensure that the resonance of chains I. and II. (III. and IV.) coincide, while the symmetry $\textbf{c}\rightarrow -\textbf{c}$ that I. and III. (II. and IV.) would rotate in the opposite sense
 for magnetic fields in the (\textbf{a},\textbf{c}) plane. A significant magnetic interaction between first neighbor chains that couples the modes would inhibit mode crossings: for weakly interacting chains the modes split first near the crossing points.

To account for quasi-linear crossing of the counterrotating modes of the antiferromagnetic regions we suggest that the interaction between chains is small along both the \textbf{b} and \textbf{c} directions. It cannot be zero, otherwise there would be no magnetic order but it must be small to explain the lack of an observable splitting. The gradual development of the magnetic order in a large temperature range suggests that the coupling between chains is due to a small concentration of defects. A small part of the sample remains paramagnetic down to 4~K.  

  The electronic overlap in the \textbf{c} direction is very small as $\text{AsF}_6$ ions separate the first neighbor chains.  On the other hand, the lack of magnetic interactions between the first neighbor chains in the \textbf{b} direction (e.g. between I. and II.) is unexpected. 
In view of the crystal and electronic structures it is surprising that the two AFMR modes are well described by Eqs.~\ref{eq1} and \ref{eq2} even near mode-crossing directions (Fig.~\ref{Fig.AFMR.ac:zoom}) as
AFMR in these high symmetry directions is rather sensitive to interactions between layers. 
The modes of interacting non-equivalent layers deviate strongly from the modes of isolated layers around regions of mode degeneracy. 
For example, the splitting of the AFMR modes of adjacent layers is well observable in the organic quasi-2D antiferromagnet, $\kappa$-(BEDT-TTF)$_2$Cu(N[CN]$_2$)Cl, despite the several orders of magnitude difference between inter- and intra-layer magnetic interactions \cite{Antal2009,Antal2011}.

\subsection{Justification of the isosceles triangular AF magnetism model} \label{justification}

The triangular magnetic structure in Fig.~\ref{Fig.Structure1}~(C) explains the magnetic frustration causing the lack of interaction between the two AFMR modes. It is the simplest frustrated order compatible with the crystal and charge density wave structure and the static magnetic susceptibility. Here we summarize the arguments supporting the trianular magnetic structure.

The quasi 1D antiferromagnetic chain behavior of molecular chains along the \textbf{a} direction follows from the known crystal structure and the high temperature magnetic susceptibility. NMR shows that highly and poorly charged molecules alternate along \textbf{a} and that the charge on poor molecules is almost one electron smaller than on rich ones  \cite{Heuze2003,Auban2009}. The full CDW structure was determined by XRD \cite{Zorina2009}.
The magnetic susceptibility at high temperatures shows that along \textbf{a} the chains are quasi 1D antiferromagnetic with a large exchange interaction between molecules. It is a natural suggestion that the ordered magnetic structure consists of these antiferromagnetic chains. From the angular dependence of the two AFMR modes we find that magnetic interactions between \textbf{a} chains running parallel in the (\textbf{a},\textbf{b}) plane are negligibly small in the antiferromagnetic state. This is explained by magnetic frustration. The negligible magnetic interaction between chains in the \textbf{b} direction does not follow from the crystal structure alone. According to band calculations the overlap between molecules in the \textbf{b} direction is not very small, the overlap integral between adjacent chains in the (\textbf{a},\textbf{b}) plane is only an order of magnitude smaller than along the chain. (Overlap in the \textbf{c} direction is very small.) Thus we propose that frustration is the reason for the weak magnetic interaction. The simplest way to obtain frustration is evident from the crystal structure. Since neighboring chains of uniformly spaced molecules are shifted by a half lattice constant, the molecular lattice is a triangular network in the (\textbf{a},\textbf{b}) plane. (Fig.~\ref{Fig.Structure1}~(B)). The uniform spacing between molecules along the chains that allows the triangular molecular structure is a unique feature of this compound \cite{Heuze2003}. To explain the magnetic frustration between neighboring \textbf{a} chains, we propose that the magnetic structure follows the “triangular” crystal structure. The main assumptions are that in the ordered state all \textbf{a} chains are simple two sublattice antiferromagnets and anisotropic exchange or other interactions between chains are small. 

\subsection{Comparison with the 2D magnetic lattice model} \label{comparison}

The two-dimensional magnetic lattice in Fig.~\ref{Fig.Structure1}~(C) serves as a model for the (\textbf{a},\textbf{b}*) plane of EDT$_2$AsF$_6$.
The isotropic antiferromagnetic interactions between first, second and third neighbor molecular pairs are characterized by $J$, $J_2$ and $J_3$ respectively.
For $J=J_2$ and $J_3=0$ the model corresponds to the regular triangular frustrated spin system.
On the other hand, if $J\gg J_2$, a two-sublattice antiferromagnetic order is established along the chains (the isosceles triangular case).
Frustration is still present in this latter case since the molecular next neighbor interactions ($J_2$) cancel and the coupling between neighboring antiferromagnetic chains vanishes.
A finite $J_3$ couples second neighbor chains and the system consists of two non-interacting 2D antiferromagnets.

The real material EDT$_2$AsF$_6$ differs somewhat from the model.
The structural transition at 190~K modifies in a subtle way the picture.
Although in the orthorhombic structure isotropic antiferromagnetic interactions between neighboring chains fully cancel, the frustration is slightly weaker in the monoclinic structure below 190~K where molecules are inclined by a few degree and the $J_2$ interactions do not cancel completely on the neighboring molecules.
The small anisotropic interactions determine the angular dependence of AFMR modes and have a profound effect on the magnetic order.

We assumed in the above argument that the chains are antiferromagnetic.
From a general point of view, it is not evident whether a $1/4$ filled chain is antiferromagnetic or ferromagnetic.
The extended Hubbard model allows for both types \cite{Mila1993}, depending on the parameters assumed in the calculations.
However, in EDT$_2$AsF$_6$ the AFMR mode diagram is incompatible with ferromagnetic chains since these would strongly interact in the (\textbf{a},\textbf{b}*) plane.
There would be no frustration in this case, independently of the sign of J$_2$ (i.e. whether the interaction between ``\textbf{a}'' chains is ferromagnetic or antiferromagnetic). 
Thus ferromagnetic chains along \textbf{a} are incompatible with the two independent AFMR modes in the  (\textbf{a},\textbf{c}) plane. 

Finally, we suggest that the N\'eel temperature of the phase transition in a perfect crystal is below 4~K. 
We explain by a variation of defect centration that the onset temperature of an observable magnetic order varies from crystal to crystal.
Defects magnetically connecting chains in the $b$ direction locally break frustration and induce an incomplete static magnetic order in regions where the average defect distance is comparable to the in-chain correlation length.  
As a result, most but not all of the crystal is antiferromagnetically ordered at 4~K.
The persistence of intrinsic paramagnetic regions signifies that at 4~K in some regions frustration overcomes the coupling between chains due to defects and residual interactions between chains. 
In these regions the same paramagnetic behaviour is observed in the ESR as at high temperatures, except for some line broadening. 
We suggest that the paramagnetic ESR corresponds to regions with small defect concentration where order is prevented by frustration at 4~K. 
The chains along \textbf{a} are quasi one-dimensional with a magnetic ordering temperature below 4~K. 
The continuous development of order and the persistance of paramagnetic regions show that the antiferromagnetic ordering temperature of the ideal, defect free system is lower than $T_N^\mathrm{NMR}= 8.5$~K suggested in Ref.~\onlinecite{Auban2009}.

Hysteresis in high fields, which is unusual in antiferromagnets, suggests also that defects strongly modify the magnetic texture of the sample, creating antiferromagnetic and paramagnetic domains.
Hysteresis due to disorder by structural defects is most important when anisotropic interactions between chains are particularly weak; this may be the case for magnetic fields in the (\textbf{a},\textbf{b}*) plane.
In organic spin-$1/2$ magnets, where there is no single ion anisotropy and anisotropic exchange between molecules composed of light elements is small, the anisotropy is mainly due to dipolar interactions.
It is a simple matter to show that in EDT$_2$AsF$_6$ the dipolar interactions between first neighbor chains cancel above the spin flop transition if the external field is in the (\textbf{a},\textbf{b}*) plane.
The fields of interest for the AFMR are well above the spin flop field; thus order is not established by dipolar interactions in the (\textbf{a},\textbf{b}*) plane.

Symmetry arguments show that in EDT$_2$AsF$_6$ the DM interaction is ineffective along  ``\textbf{a}'' if  chains are antiferromagnetic.
On the other hand, the DM interaction is not zero between neighbor ``\textbf{a}'' chains in the  (\textbf{a},\textbf{c}) plane. A small, frequency dependent broadening of the ESR line has been attributed to the DM interaction between chains in the paramagnetic state \cite{Nafradi2010}. A negligible ferrimagnetism is expected in the magnetically ordered regions, since the chains are antiferromagnetic with a large exchange, J in the ordered state. This is in accord with the absence of weak ferromagnetism in the static magnetization. However, the DM interaction can influence the magnetic order in a magnetic field dependent way. The relative orientation of sublattice magnetizations in neighboring chains can change abruptly when the external field direction is swept through the DM vector. The DM interaction may be at the origin of the hysteresis of the AFMR modes. Indeed, the DM vector of interchain interaction lies in the (\textbf{a},\textbf{b}) plane and hysteresis effects are most pronounced when the magnetic field is rotated in this plane.

\section{Conclusion}
In conclusion, EDT$_2$AsF$_6$ is a quasi-one-dimensional compound which has a 1/4 filled electronic band with unusual electronic and magnetic properties.
We find that the (\textbf{a},\textbf{b}*) plane of EDT$_2$AsF$_6$ is to a good approximation a 2D frustrated isosceles triangular magnetic lattice with a strong isotropic exchange interaction, $J$ on the base and a much weaker interaction, $J_2$ on the legs.
Frustration remains important in this triangular lattice; the ideal system decomposes into two networks of non-interacting one-dimensional antiferromagnetic chains preserving an overall paramagnetic response down to  temperatures as low as 4~K.
The AFMR modes show that the magnetic interaction between closely lying chains in the (\textbf{a},\textbf{b}*) plane is unexpectedly weak due to frustration.
Structural defects lift the frustration and nucleate antiferromagnetic regions up to temperatures as high as 40~K.

\begin{acknowledgments}
Support from the projects T\'AMOP-4.2.1/B-09/1/KMR-2010-0002, OTKA K107228.
The work in Lausanne was supported by SCIEX by the Swiss National Science Foundation (Grant No. 200021\_144419) and ERC advanced grant ``PICOPROP'' (Grant No. 670918).
The work in Angers was supported by the ANR grant 3/4-Filled 2009-2011 (ANR-08-BLAN-0140-01) and the CNRS.
\end{acknowledgments}

\bibliography{./EDT_AFMR_01}

\end{document}